\documentclass[useAMS,usenatbib,usegraphicx]{mn2e}

\newcommand{\msun}{M$_\odot$}
\newcommand{\lsun}{L$_\odot$}
\newcommand{\mic}{$\mu$m}

\title[Missing SNe at high redshift]
{How many supernovae are we missing at high redshift?}

\author[Mannucci, Della Valle \& Panagia]{
F. Mannucci$^1$\thanks{E-mail:filippo@arcetri.astro.it}
M. Della Valle$^2$, and
N. Panagia$^{3,4,5}$\\
$^1$INAF - Istituto di Radioastronomia, Largo E. Fermi 5, 50125 Firenze, Italia\\
$^2$INAF - Osservatorio Astrofisico di Arcetri,Largo E. Fermi 5, 50125  Firenze, Italia\\
$^3$STScI, 3700 San Martin Drive, Baltimore, MD 21218, USA\\
$^4$INAF-HQ,  Via del Parco Mellini 84, I-00136 Rome, Italy\\
$^5$Supernova Ltd., Olde Yard Village \#131, Northsound Road, Virgin Gorda,
British Virgin Islands
}

\begin{document}
\date{Submitted; Accepted}
\pagerange{\pageref{firstpage}--\pageref{lastpage}} \pubyear{2007}
\maketitle

\begin{abstract}
Near-infrared and radio searches for core-collapse supernovae
(CC SNe) in the local universe have shown
that the vast majority of the events occurring in massive starburst
are missed by the current optical searches as they explode in
very dusty environments. 
Recent infrared observations have shown that the fraction of 
star-formation activity that takes place in very luminous 
dusty starbursts sharply increases with redshift and becomes the dominant 
star formation component at z$\ge$0.5. 
As a consequence, an increasing fraction of SNe are expected to
be missed by high-redshift optical searches. 
We estimate that 
5--10\% of the local CC SNe are out of reach of the
optical searches.
The fraction of missing events rises sharply toward z=1, when about 
30\% of the CC SNe will be undetected. At z=2 the missing 
fraction will be about 60\%. 
Correspondingly, for type Ia SNe, our computations provide 
missing fractions of 15\% at z=1 and 35\% at z=2.
Such large corrections are crucially important to compare
the observed SN rate with the expectations from the
evolution of the cosmic star formation history, 
and to design the future SN searches at high redshifts.
\end{abstract}

\begin{keywords}
supernovae:general --- galaxies:starburst
\end{keywords}

\section{Introduction}
\label{sec:intro}

\subsection{Infrared galaxies}

Most of the star formation activity in the local universe occurs in ``normal''
galaxies characterized by star formation rates (SFRs) of 0.1--5 \msun/yr and 
moderate dust extinctions ($A_V<1$). Nevertheless, a few objects with 
high SFRs (100-1000 \msun/yr) 
and heavy dust attenuation ($A_V$=5--50 mag) exists. 
Only a very small fraction of the UV light produced by their young stars
can escape these galaxies, while most of the energy is converted into thermal
infrared (IR) radiation by heating the dust 
(e.g., Burgarella et al. 2006; see also Mannucci \& Beckwith, 1995).
As a consequence their bolometric luminosity 
is dominated by the far-IR radiation from the hot dust.
These galaxies are classified as Luminous IR Galaxies (LIRGs) if their 
total far-IR luminosity is above $10^{11}$~\lsun\ 
(corresponding to SFR$\sim$10--20 \msun/yr, Choi et al. 2006)
and as Ultraluminous IR galaxies (ULIRGs) above $10^{12}$ \lsun. 
Most LIRGs and ULIRGs show the signs of recent major mergers 
(e.g., Lagache et al. 2005) 
that are considered to be the origin of the strong episode of star formation.
Liang et al. (2004), Marcillac et al. (2006) and Choi et al. (2006)
have studied samples of distant (z$\sim$0.8) LIRGs 
and found no evolution of the properties of these
galaxies with redshift. 

Even if these objects are very active in forming stars, 
their comoving volume density is low in the local universe
and their contribution to the total star formation 
activity is below 5\% (Soifer \& Neugebauer, 1991).
Two decades ago, using data from the IRAS satellite, it was discovered that
massive starburst galaxies were much more numerous and/or luminous 
in the past than they are today. 
For example, Hacking et al. (1987) have shown that the counts 
of the bright IRAS galaxies cannot be explained without a fast 
evolution up to z=0.2.
Using the Spitzer satellite, the evidence for rapid evolution
has been recently extended up to z=1 (Le Floch et al. 2005) 
and beyond  (P\'erez-Gonz\'alez et al. 2005; Daddi et al. 2005).
As a consequence, it is now well established that massive dusty starburst 
are rare objects in the local universe but were much more 
common a few Gyr 
ago and dominated the star formation (SF) activity before z=2.\\

\subsection{Core-Collapse supernovae}

In the last few years, another class of astronomical objects, 
supernovae, (SNe) has proved to
be very useful in observational cosmology. Besides the well known results
about the accelerated expansion of the universe obtained 
from studies of type Ia SNe,
also CC SNe in high-redshift galaxies have attracted much interest. 
These objects, which are generated by the gravitational collapse of
very massive stars (M$>$8\msun), 
are among the most important producers of metals 
(e.g., Matteucci \& Greggio, 1986),
are considered to be responsible for dust production
in the young universe
(Maiolino et al. 2004a, 2004b), could dominate the feedback 
in the process of galaxy formation (see, for example, Ferrara \& Ricotti,
2006, and references therein), and can be used to estimate the SF density
at high redshifts (e.g., Dahlen et al. 2004).

The measured rates of CC SNe both at low and at high redshifts are based on
optical observations because only at these wavelengths the current
instrumentation has sufficient field coverage, 
spatial resolution and sensitivity
to detect large numbers of SNe within a reasonable observing time.
In the local universe (z$<$0.1), the best available 
rates have been computed by Cappellaro et al. (1999) 
and Mannucci et al.  (2005) and are based
on a compilation of a few visual and photographic searches. Other
searches for CC SNe based on CCD imaging are available or currently
active (Filippenko et al. 2001; Strolger, 2003). 
Recently, the rates for higher redshifts (z$\sim$0.5) CC SNe 
have been measured 
by Dahlen et al. (2004) and Cappellaro et al. (2005), although
large uncertainties remain, for example because of the lack of an 
extensive spectral identification of the SNe candidates.
To derive the intrinsic SN rates from the observed number of events
it is necessary to estimate the fraction of SNe that are missed 
because of any known reason, 
as dust extinction in the host galaxies and the reduction of
detection efficiency near the galaxy nucleus. 
Customarily, these corrections are estimated
in the local universe and are assumed to hold unchanged with 
redshift. Also, they are appropriate to correct the SN rates 
in normal galaxies and do not take into account the presence of LIRG
and ULIRG among the galaxy target. For
example, Dahlen et al. (2004) assume a moderate average extinction
of E(B--V)=0.15, corresponding to $A_V\sim0.5$, appropriate for the local 
population of normal galaxies, and use this correction at all redshifts. 
The resulting rate accounts only for the SNe exploding in galaxies where 
the star formation activity occurs in relatively clean enough
environments so that significant fraction of SNe can be detected. 
SNe exploding inside dusty 
starbursts are missing from these counts.

\begin{figure*}
\includegraphics[angle=-90,width=18cm]{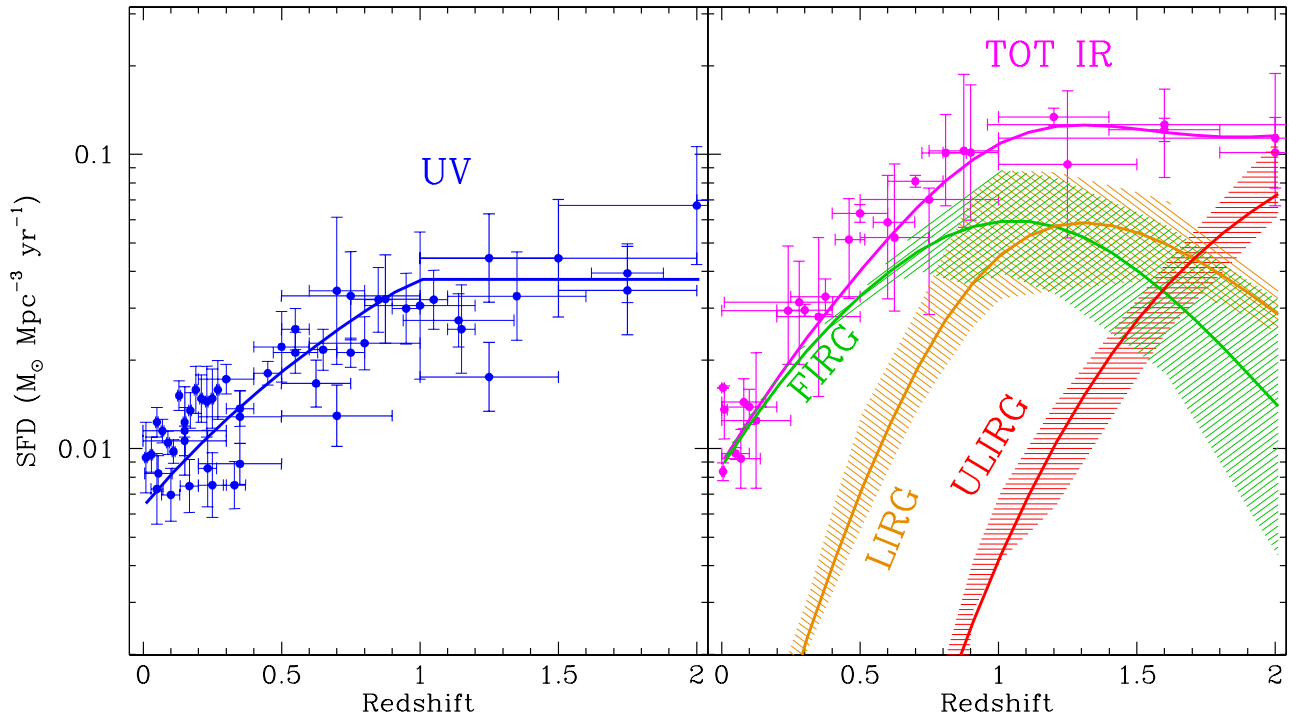}
\caption{
\label{fig1}
{\em Left panel:} Evolution of the SF density with redshift as 
derived from UV observations, uncorrected for dust extinction. 
Data points are from
Lilly et al. (1996), Connolly et al. (1997), Treyer et al. (1998),
Cowie et al. (1999), Sullivan et al. (2000), Massarotti et al. (2001), 
Wilson et al. (2002), Wolf et al. (2003),  Schiminovich et al. (2005),  
Baldry et al. (2005), Wyder et al. (2005), Thompson et al. (2006),
and Dahlen et al. (2006). 
Original data have been first converted 
into luminosity density at 2800\AA\ using a UV slope $\beta=-1.5$,
and then to a SF density as in Baldry et al. (2005).
The solid line is the fit to these data points in eq.~\ref{eq:uv}.
{\em Right panel:} Evolution of the SF density with redshift as 
derived from IR and radio observations. The data points are from
Condon (1989), Flores et al. (1999), Barger et al. (2000), 
Haarsma et al. (2000), Machalski \& Godlowski (2000), Sadler et al. (2002), 
Serjeant et al. (2002), Condon et al. (2002), 
Georgakakis et al. (2003), P\'erez-Gonz\'alez et al. (2005),
Mauch (2005) and Babbedge et al. (2006).
The dashed regions in green, orange and red show the range of possible 
contributions from FIRG, LIRG and ULIRG, respectively, 
as derived by P\'erez-Gonz\'alez et al. (2005). The thick lines in the same 
colors are the assumed evolution of the three populations 
as described by eqs. \ref{eq:firg}, \ref{eq:lirg}, and \ref{eq:ulirg}.
The magenta line is the total SF density given by the sum of the 
three contributions, and can be directly compared with the data points.
}
\end{figure*}

\subsection{Supernova searches at longer wavelengths}

In the local universe, optical searches for SNe in LIRGs and other
starburst galaxies (Richmond et al. 1998)
failed to detect the expected enhancement of the rate
over more quiescent galaxies
due to the strong activity of star formation.
To test the hypothesis that this shortage is due to the presence of large
amounts of dust obscuring a substantial fraction of SNe, several groups 
(van Buren \& Norman, 1989; 
Maiolino et al. 2002; Mannucci et al. 2003; Mattila et al. 2004; 
Cresci et al. 2006) 
have carried out near-IR searches for SNe in starburst galaxies.
As dust extinction is about 10 times less important in the K-band than 
in the optical, these searches were expected to detected substantially
higher rates. Indeed, some of these searches 
have detected  a population of near-IR CC SNe that are missed by
optical searches. For example, 2 SNe
(SN1999gw, Cresci et al. 2002, and SN2001db, Maiolino et al. 2001)
out of the 4 events detected in the K-band 
by Maiolino et al. (2002) are missing from the optical catalogs. 
At least one of them is characterized by 5--6 magnitudes of 
extinction in the V band
which would have prevented its detection at optical wavelengths.
However, these searches failed to 
recover the majority of the expected events. 
Mannucci et al. (2003) demonstrated that about 80\% of the CC SNe expected to
explode in LIRG and ULIRG, on the basis of their far-IR luminosity,
are still undetected even by near-IR searches.
Similar results have been obtained by Cresci et al. (2006) for 
a SN search with the near-IR camera NICMOS on the HST.
The shortage of SNe can be due either to the presence of very large 
dust column densities ($A_V>20$) or to the fact that most of the SNe
explode in the nucleus of the starburst galaxy, where 
SNe cannot be easily disentangled from the bright emission of
the host galaxy.
In both cases, most of these SNe are missing from the
estimates of the local rates, and their rate cannot be inferred 
from the SNe detected in normal galaxies.
Recently, Lonsdale et al. (2006) have monitored the nucleus of Arp~220,
the prototypical ULIRG, at radio wavelengths with sub-arcsec resolution. 
They detected a few faint variable sources and a 
population of brighter, slowly variable point sources. 
If these data are interpreted in terms of
SNe and SN remnants, the number of SN events would be enough 
to power all the far-IR luminosity of Arp~220. However, none of this
object could be detected by optical or near-IR searches as they are
deeply embedded in the huge amount of nuclear dust.

The effect of dust extinction over the observed SN rate is expected to be
severe for LIRG and ULIRG.
Many authors, such as Cram et al. (1998), Jansen et al. (2001),
Hopkins et al. (2001), Sullivan et al. (2001),
Kewley et al.  (2002), P\'erez-Gonz\'alez et al. (2003) and
Choi et al. (2006), have studied the amount of extinction 
in starburst galaxies as a function of the SFR,
and concluded that the two quantities are strongly correlated, i.e., 
the more active galaxies also show higher levels of dust extinction.
As a consequence, both the SFR and the SN rate can be derived by optical and
UV observations for galaxies with SFR$\le$5\msun/yr because the low average
extinction ($A_V\le1$) allows for a meaningful correction. 
On the contrary, for larger SFRs, the average extinction becomes
very large. For example 
Choi et al. (2006) have studied a sample of LIRG
at z$\sim$0.8 and have obtained an average  extinction of $A_V=2.5$, which
corresponds to an average reduction of the flux in the [OII]$\lambda3727$ line
of 97\%.\\
For these galaxies the total SF activity can only
be derived at wavelengths less sensitive to dust extinction
(far-IR and radio), while the total SN rate cannot be measured. \\

In this paper we want to estimate the fraction of SNe which are missed
both in the local and in the distant universe 
by current and future optical or near-IR SN searches.
In the next section we will review the information on the SF
activity at z$<$2 as derived from UV and far-IR observations
in order to quantify the star-formation activity occurring in clean and dusty
environments. In section~\ref{sec:snrate} we will use the resulting SF density
to derive the SN rate. 
Throughout this paper we will use the concordance cosmology 
$(h_{100},\Omega_m,\Omega_{\Lambda})=(0.7,0.3,0.7)$.

\section {Evolution of the luminous infrared galaxies}

In the last decade, many authors have contributed to plot a coherent picture 
of the evolution of the star formation activity along the Hubble time 
(see Hopkins \& Beacom 2006 for a recent review). 
Observations at radio, IR, optical, UV and X-ray wavelengths have
detected the SF activity over a broad range of redshifts. The
emerging picture shows a fast increase of the SF density with increasing
redshift between 0 and 1, a roughly constant density between z$\sim$1 and
z$\sim$4, and a possible decrease for z$>$4
(Bouwens \& Illingworth 2006, Mannucci et al. 2007).
Observations made at different wavelengths
sample different aspects of the star formation
activity. The UV observations are deeply affected by the presence of dust, 
and evidence of significant extinction is found in practically 
all types of 
high-redshift star-forming galaxies. On the contrary, far-IR 
observations are more suitable to trace the star formation that takes place in
dusty environments. It is now clear that a complete census of the star
formation can be obtained only by putting together all these contributions.\\

P\'erez-Gonz\'alez et al. (2005) have used Spitzer data at 24\mic\ of about
8000 galaxies with reliable spectroscopic or photometric redshifts 
to constraint
the evolution of the far-IR Luminosity function (LF) with redshift. 
They confirmed the strong evolution of the far-IR luminosity density with
redshift and, most importantly, were able to trace the different contributions
to this evolution (see Fig.~1). At z$<$1, the dominant contribution comes
from galaxies with low IR luminosities 
(faint IR galaxies, FIRGs, L$<10^{11}$\lsun). The LIRG component accounts
for a few percent of the total SF density in the local universe but has a
faster evolution with redshift, becoming the dominant component at z$\sim$1.
At higher redshifts, the LIRG contribution declines and the SF density is
dominated by ULIRGs.

Several effects contribute to the uncertainties of this picture, especially
at z$>$1:
the value of the faint-end slope $\alpha$ of the LF is poorly constrained;
the total far-IR luminosity in the 8--1000\mic\ wavelength range 
is estimated from the flux at one single wavelength; 
most of the redshift determinations are photometric, 
allowing the presence of a significant fraction of interlopers.
Despite all these uncertainties, the fast evolution up to z=1 is 
a strong results confirmed by many studies, as explained in the introduction. 
For example, using a smaller sample of galaxies but with a larger 
fraction of spectroscopic redshifts, 
Le Floch et al. (2005) found the same type of evolution for FIRG, LIRG and
ULIRG up to z=1, even if they found a larger total number of LIRG.
Also Chary et al. (2007) find the same behavior, with a ``FIRG epoch'' 
at z$<$1, a LIRG epoch at 1$<$z$<$2, and a ULIRG epoch at z$>$2.
The most uncertain effect is the reduction in number of LIRG at z$>$1 and the
corresponding increase in the ULIRG comoving density, but
this behavior has hardly any effect on the results of this paper because the
fractions of SNe detected in LIRG and ULIRG are very similar (see next
section).

Fig.~1 shows the observed evolution of the SF density with  redshift as
measured in the UV and in the far-IR. Data points are derived from the
sources listed in the figure caption and assume the ``SalA'' 
modified Salpeter initial mass function (IMF, Baldry \& Glazebrook 2003, 
see also Hopkins \& Beacom 2006). UV data are uncorrected for dust 
extinction.
The dashed areas in the IR panel show the uncertainties
of the various contributions from P\'erez-Gonz\'alez et al. (2005).
The solid lines show the SFD 
deriving from the observed UV galaxies, FIRG, LIRG and ULIRG. 
To describe these curves in an analytic form, we use a 
parametrization in which a power-law of (1+z) is multiplied
by an exponential cutoff at high redshifts:

\begin{eqnarray}
\log (SF_{FIRG}) & = & -2.05+3.5\log (1+z) \nonumber         \\
                 &   & -\log\left[1+\exp(3.7\ (z-1.1)\right]
\label{eq:firg}
\end{eqnarray}
\begin{eqnarray}
\log (SF_{LIRG}) & = &-3.65+8.8\log (1+z)    \nonumber \\
                 &   &-\log\left[1+\exp(4.6\ (z-0.96)\right]
\label{eq:lirg}
\end{eqnarray}
\begin{eqnarray}
\log (SF_{ULIRG}) & = &-5.90+12.5\log (1+z)    \nonumber \\
                 &   &-\log\left[1+\exp(3.0\ (z-1.1)\right]
\label{eq:ulirg}
\end{eqnarray}
 \begin{equation}
 \log (SF_{UV})= \left\{
     \begin{array}{ll}
      -2.20+2.6\log (1+z)~~~~   & \rm{if}~~ z\le 1 \\
      -1.43                    & \rm{if}~~ z>1 
	  \end{array}
 \right.
\label{eq:uv}
\end{equation}

These analytic equations are valid up to z=2.
Since these contributions to the cosmic SF activity occurs in
different environments, each of them is associated to a different {\em
fraction} of SNe that can be detected.
As a consequence, while the evolution of the 
{\em intrinsic} CC SN rate follows the total SF density,
the {\em observed} SN rate is expected to 
show a different behavior, because the fraction of detectable SN changes with
redshift. 

\begin{figure*}
\includegraphics[angle=-90,width=18cm]{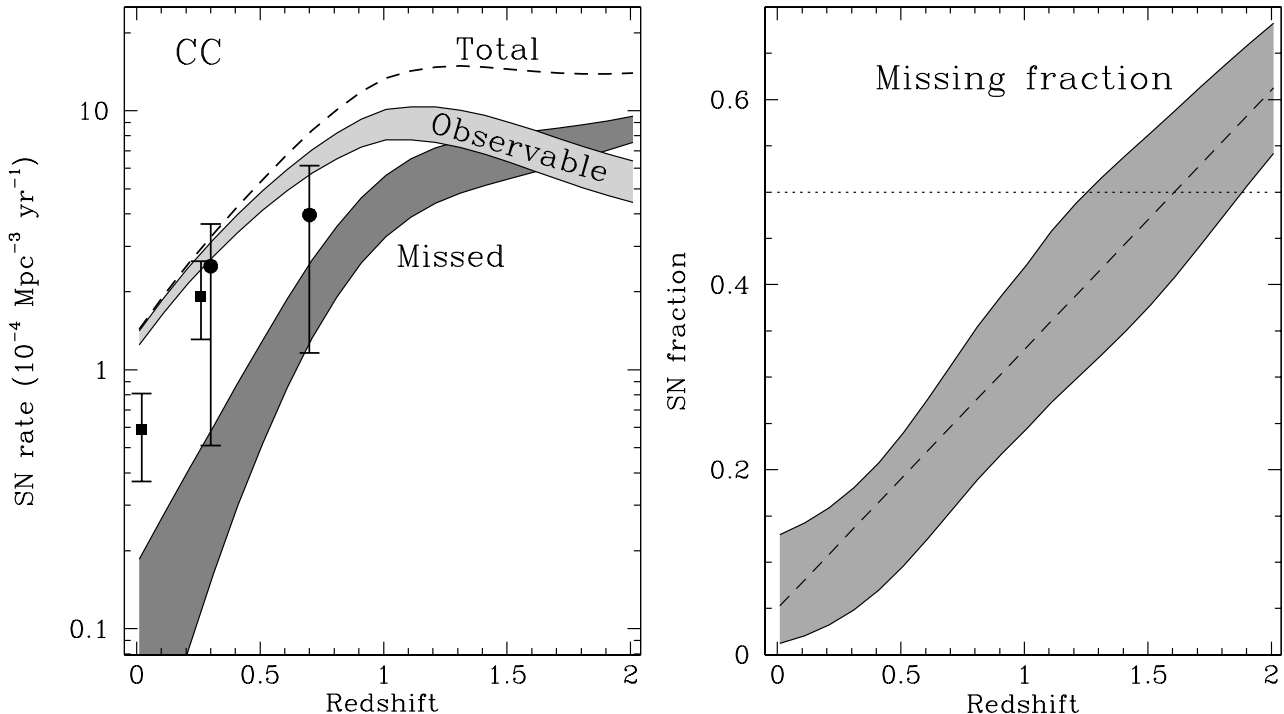}
\caption{
\label{fig2}
{\em Left panel:} Evolution of the CC SN rate with redshift. 
The dashed line shows the intrinsic total SN rate derived from the SFD.
The SN rate that can be recovered by optical searches
is shown in light grey, and is compared with the data points
from Cappellaro et al. (2005) (squares),
and Dahlen et al. (2004) (circles). For consistency, the former two data 
points have been converted to comoving rates as in the latter work.
In dark grey we show the expected SN rate for SNe that cannot be detected 
because they are exploding in very dusty starburst, as explained in the text.
The reported uncertainties correspond to the possible range 
of the fraction of SNe that can be detected in the various classes of
galaxies, as explained in Sect.~\ref{sec:snrate}.
{\em Right panel:} Fraction of CC SNe than are not present in the
optical and near-IR searches as a function of redshift.
The dashed line shows the results of eq.~\ref{eq:fracCC}.
}
\end{figure*}

\section {Intrinsic and observed SN rates}
\label{sec:snrate}

\subsection{Core-Collapse Supernovae}
\label{sec:ccrate}

The {\em total} CC SN rate at any redshift is proportional to the total
SFD at that epoch as the lifetime of the exploding
stars is short, of the order of $10^7$ yr, and can be neglected. 
The proportionality between SN rate and SFD depends on the assumed IMF.
If $\rho$ is the SN rate in units of SN/yr and $\psi$ is the SFR in
units of \msun/yr, a simple integration of the Salpeter IMF between 8 and 50
\msun\ provides $\rho=0.007\psi$ (see, for example, Cappellaro et al. 2005).
For the SalA IMF, Hopkins \& Beacomp (2006) obtain  $\rho=0.00915\psi$.
Nevertheless, it should be noted that 
{\em the total SN rate expected from the UV and far-IR observations
is practically independent of the assumed IMF},
as both the UV and far-IR luminosities are mostly due to the same 
massive stars producing the CC SNe.  
For example, the use of a Salpeter instead of a SalA IMF 
would result in a SFD 30\% higher than that shown in Fig.~1, 
but the expected
number of SNe for a given SFR would be 30\% lower, and the two factors
would cancel out each other.
On the contrary, the stellar mass range to produce a CC SNe is an
important parameter which is only weakly constrained by models. Here we use
the ``canonical'' range between 8 and 50 \msun 
(e.g., Woosley \& Weaver, 1986), 
while the use of different mass ranges would
produce different total number of CC SNe. For example, using a range
between 10 and 40 \msun\ would reduce the expected rate of 32\%.
The total CC SN rate we derive is shown in Fig.~2.
It is very similar to that in Hopkins \& Beacom (2006) 
apart from the different parametrization used for the SFD.\\

To  derive the {\em observed} SN rate from the measured SF density 
we need to add a few assumptions
about the average properties of the galaxies of the different
populations: 
\begin{enumerate}
\item 
Since the star formation activity detected in UV selected galaxies
and in FIRG
occurs in  relatively clean environments, we assume that the optical SN 
searches are adequate to measure the rates. This means that many CC SNe in
these galaxies are detected, and that the missing part can be recovered
by estimating the average dust extinction as in Cappellaro et al. (1999)
or in Dahlen et al. (2004).  This is justified by the fact that 
low values of $A_V$ are usually found in these galaxies. 
Also at high-redshift, 
most of the UV-selected Lyman Break Galaxies (LBG) show $A_V<1$ 
(Vijh et al.  2003; Burgarella et al. 2006). 
For these reason, we assumed that all the SNe
in UV-selected galaxies and at least 80\% of those in FIRG 
are accounted for by the optical searches,
either directly or through the extinction corrections.
\item In the local universe, less that 10\% of the SNe in LIRG are detected 
by optical searches, and less then 20\% can be recovered by near-IR searches
(Mannucci et al. 2003).
At higher redshifts, 
the fraction of detected SNe could be higher if distant LIRG have a 
much smaller content of dust, for example due to a lower metallicity.
To allow for any possible, although unobserved, 
evolution in the dust properties of LIRG, we assume that the 
{\em maximum} fraction
of SNe that are accounted for by the optical searches is 0.2 at z=0 and
linearly increases to 0.4 at z=2. This is a solid upper limit as 
optical searches are not likely to detect 
more that 40\% of the SNe in LIRG at any redshift.
The {\em minimum} fraction of detectable SNe is assumed to be 0 at any redshift.
\item Only a minor fraction of SNe are detected in the ULIRG in the local
universe, as these galaxies are known to enshroud their star formation
activity inside large amounts of dust. 
For example, Arp~220 is expected to produce 4 SNe per year (Mannucci et al.
2003), while only one possible SN was ever detected (Cresci et al. 2006), and 
only by using the HST near-IR camera.
As a consequence, we assume that
between 0 and 10\% of the SNe in ULIRG are present in the rates derived from
optical/near-IR observations.
\end{enumerate}
All these quantities are rather uncertain, but the assumed minimum and maximum
fractions are expected to encompass all the possible values.

By using these assumptions and the SFDs described in 
eqs.~\ref{eq:firg}--\ref{eq:uv} we derive the fraction of SNe 
that can be recovered by optical searches.
The results are shown in Fig.~2 where the expectations are compared with the
current set of observations. In the local universe only a small fraction
of SNe (5--10\%) are not accounted for by the optical rates, a fraction
which is smaller that the uncertainties in the measured rate. Going toward
higher redshifts, this fraction increases rapidly, reflecting the rapid
evolution of the number density of LIRGs and ULIRGs. 
At z=1 the undetected fraction 
is 20--40\%, rising to $\sim$60\% at z=2.

The behavior of the missing fraction $f$(CC) of CC SNe can be described by a linear
relation with redshift: 
\begin{equation}
f(\rm{CC})=0.05+0.28\ \rm{z}~~~~~~~~\rm{for~~z}\le2
\label{eq:fracCC}
\end{equation}

\begin{figure*}
\includegraphics[angle=-90,width=18cm]{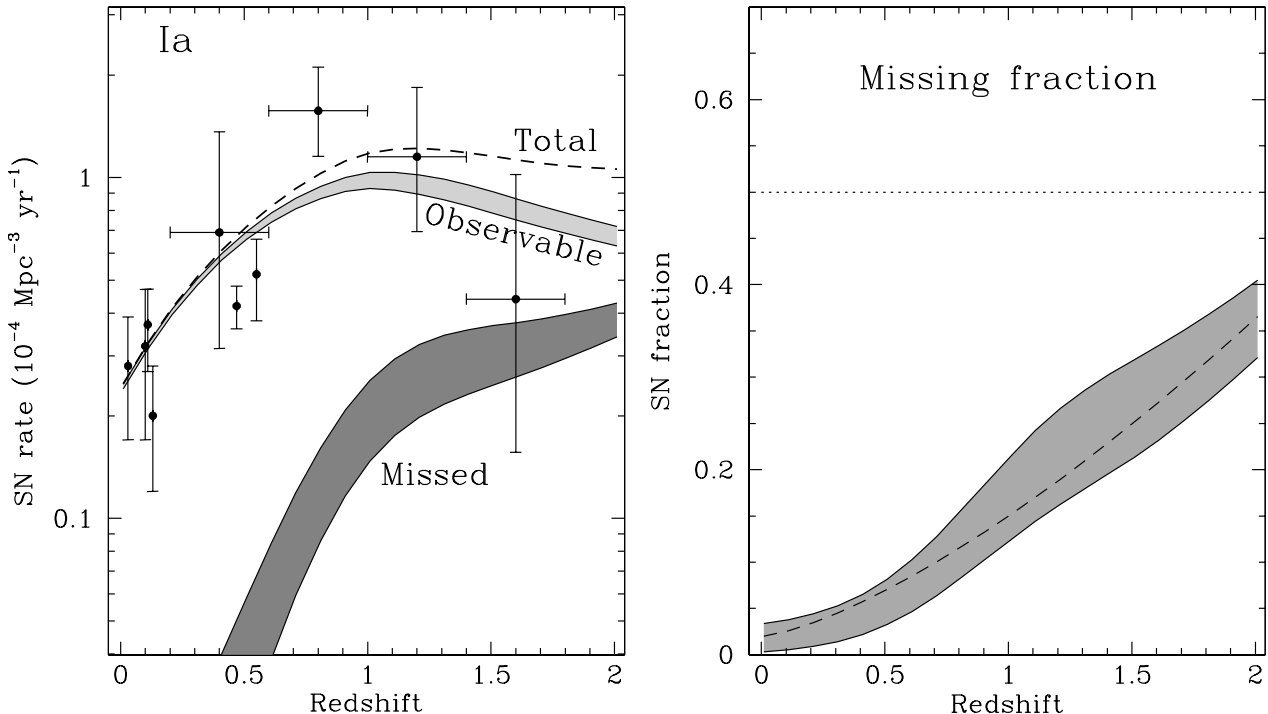}
\caption{
\label{fig3}
{\em Left panel:} Evolution of the type Ia SN rate with redshift. 
See the caption of Fig.~2 for the explanation of the symbols.
The observed rates, shown by the dots, are from
Mannucci et al. (2005), Magdwick et al. (2003), Strolger (2003), Blanc et al.
(2004), Pain et al. (2002), Neill et al. (2006) and Dahlen et al. (2004).
{\em Right panel:} Fraction of type Ia SNe than are not present in the
optical and near-IR searches as a function of redshift.
The dashed line shows the results of eq.~\ref{eq:fracIa}.
}
\end{figure*}

\subsection{Type Ia supernovae}
\label{sec:iarate}

Type Ia SNe are also affected by dust extinction as they explode in all
environments. For example, in the local universe there are
type Ia SNe that show large
amounts of dust extinction, such as 
SN2002cv (Di Paola et al. 2002) and 
SN2000E (Valentini et al. 2003).
Chary et al. (2005) have studied 50 Ia SN host galaxies at 0.1$<$z$<$1.7
and found larger amounts of dust than in normal field galaxies, and far-IR 
luminosities similar to those in CC SN hosts. This is an indication
that the environments of the two SN populations are similar.

While all the CC SNe are expected to explode soon after the formation of the
massive parent star, the Ia SNe show a wide delay time distribution 
(DTD; i.e., a wide distribution of the time elapsed between the
birth of the progenitor star and the SN explosion). 
According to many computations (e.g., Greggio \& Renzini 1983, Greggio 2005)
the DTD could range between 10$^7$ and 10$^{10}$ years.
Such a large spread is needed to 
reproduce the variation of the Ia SN rate per unit mass with host 
galaxy color (Mannucci et al. 2005, 2006; Greggio, 2005)
and to account for the metallicity evolution of the galaxies 
(e.g., Matteucci et al. 2006, Calura \& Matteucci 2006).
Mannucci et al. (2006) have shown that 
the dependence of the Ia SN rate in early-type galaxies with the 
radio power of the host (Della Valle et al. 2005) is best explained if
about half of the Ia SNe,
the so called {\em prompt} component,
explode within 10$^8$ yr from star formation,
while the rest (the {\em tardy} component) has a much wider DTD. 
For this reason about half of the Ia SNe (the {\em prompt} component) 
should have 
extinction properties similar to the CC SNe as their progenitors 
explode on similar timescales. 
The time evolution of the extinction of the {\em tardy} component 
is more uncertain. 
In a naive picture we can assume that all the {\em prompt}
population suffers of the same extinction of the CC SNe, as explained in 
section~\ref{sec:ccrate}, while the 
{\em tardy} population has time to escape out of
the very dusty star-forming
region and, as a consequence, show a much milder extinction and 
can easily be recovered by observations.
Figure~3 shows the results of the computations based on these hypothesis
and using Mannucci et al. (2006) DTD:
the fraction of missing Ia SNe increases from a few percent in the
local universe to about 15\% at z=1 and 35\% at z=2. This 
behavior can be described by:
\begin{equation}
f(\rm{Ia})=0.02+0.12\ \rm{z}^{1.4}~~~~~~\rm{for~~z}\le2
\label{eq:fracIa}
\end{equation}
We point out that the
results for type Ia SNe are somewhat more uncertain 
than those for CC SNe, because they also
depend on the actual DTD and on the evolution of the extinction with
delay time.\\

The expected volumetric rates in Figures~2 and 3 can be translated into
the expected number of SNe per sq. arcmin per year that can be detected by
any survey as a function of the maximum redshift reached, as plotted in
figure~4. For example, a survey capable of detecting CC SNe up to
z=2 is expected to find 3 events per year for each sq. arcmin monitored
out of the 5 CC SNe exploded.

\begin{figure}
\includegraphics[width=8.5cm]{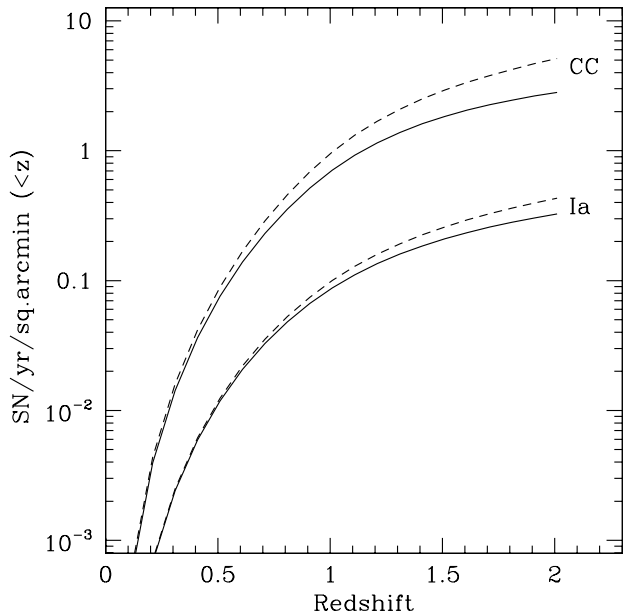}
\caption{
\label{fig4}
Predicted cumulative number of CC and Ia 
SNe exploding in one year in 1 sq. arcmin 
of sky from z=0 up to a given redshift. 
The dashed line shows the total number of exploding
SNe, the solid line the SNe that can be detected by optical and near-IR
searches.
}
\end{figure}

\section{Discussion}

\subsection{Uncertainties}
The results in Fig.~2 and Fig.~3 depend on the various assumptions, especially
the somewhat arbitrary division of the galaxies in 
FIRG, LIRG and ULIRG,
and the fraction of SNe that are assumed to be detected in each
of these classes. 
Even if some uncertainties are present, a robust conclusion
is inevitable: as most
of the the star formation activity at z$>$1 occurs in dust-rich
LIRGs and ULIRGs, optical SN searches will not follow the global
SN rate but only the fraction of detectable SNe, which will 
decrease sharply as the redshift increases.
In particular, it is impossible that the fraction of SNe missed at
high redshift is similar to that lost in the local universe 
(about 10\% for the CC SNe).
For example, such a low level of incompleteness would be possible at z=1
only if
more than 75\% of the CC SNe in LIRGs are dected by optical/near-IR searches;
and at z=1.5, not only all the CC SNe in LIRG should be detected, 
but also 40\% of those in ULIRGs. 
These high detection efficiencies would imply a strong evolution of the dust
properties of LIRGs and ULIRGs, evolution that is not observed 
(Liang et al. 2004, Marcillac et al. 2006, Choi et al. 2006).
As a consequence, although the exact amount remains uncertain, 
we conclude that at z$>$1 a large fraction of SNe remain 
unaccessible to optical/near-IR searches.

\subsection{Comparison with the observed rates for Core-Collapse SNe}

Hopkins \& Beacom (2006) have already derived the total CC SN rate
from the cosmic SFD. In their Fig.~7 they compared the expectations with the
observations, and found that in all cases the observed rates are below 
the expectations. They attribute this finding to several effects, 
such as incompleteness
of the local rate and insufficient dust correction for distant rates.
The same effect can be seen in our Fig.~2, but 
in this case the comparison must be done with the observed fraction instead of
the total one. The discrepancy is much reduced, especially at high
redshifts:
the presence of a significant fraction of SNe that cannot 
be detected helps in matching the expected and observed SN rates.

This remaining systematic overprediction of the rates can be attributed to 
several effects. 
As discussed in section~\ref{sec:snrate}, it could be due
the mass range for CC production, which could be significantly different
(narrower) than the 8--50 \msun\ range as adopted here.

Another potentially important effect is related to 
the origin of the far-IR flux in the FIRG galaxies.
It is well known that in the quiescent galaxies
a significant fraction of far-IR luminosity is not linked to the young 
stellar populations but rather to the general radiation field.  
For this reason the SFD for the FIRG in figure 1, derived from the far-IR data, 
could be overestimated because part of this far-IR flux could be related to 
old rather than young stars. As the SN counts in the local universe 
are dominated by 
this FIRG component, the actual total number of SN at low redshift could 
be lower than the predictions shown in the left panel of
figure~2, more in agreement with the measured rates. 
At variance, the {\em fraction} of missing SN in the right panel of 
figure~2 is not very sensitive to the total normalization of SN in FIRG 
as it effects both detected and undetected SNe.

It should be noted that 
the local rate from Cappellaro et al. (1999) 
($0.59\pm0.22\cdot10^{-4}$ Mpc$^{-3}$ yr$^{-1}$)
is the only value more then 1$\sigma$ below the expectations
($1.3\pm0.1\cdot10^{-4}$ Mpc$^{-3}$ yr$^{-1}$).
Indeed, this value is based on a compilation of visual and 
photographic searches that could miss a significant fraction of SNe, 
for example near galaxy nuclei. 
Cappellaro et al. (1999) considered this effect, but 
the applied correction may have been not large enough.
This fact can be understood considering
the recent results of the 
Lick Observatory SN Search (LOSS, Filippenko et al., 2001). 
The LOSS search has a
deeper limiting magnitude and, in practice, is a volume limited search
also for the CC SNe.  The fraction of CC SNe in the LOSS sample (Li, 2005)
is considerably larger than in Cappellaro et al. (1999) sample, 
indicating that the latter sample may 
miss a significant fraction of CC SNe.\\

\subsection{Comparison with the observed rates for type Ia SNe}

Many more measurements of type Ia SN rate are available (see Fig.~3)
although the spread among different results is sometimes larger than
the quoted errors, a fact that suggests the presence of uncorrected
systematic effects.
Data show a marked increase of the rate between z=0 and z=1 and, possibly,
a constant rate at higher redshifts. 
The wide DTD reproducing the rates in the local universe in the various
classes of galaxies (for a full discussion see Mannucci et al., 2006)
tends to produce more SNe at z$>$1 than observed by Dahlen et al. (2004).
Even if these observations are very uncertain, as they are based on a small
number of confirmed SNe and on several untested assumptions, the presence
of dust extinction helps in reconciling expectations with observations.

\subsection{Other sources of incompleteness}

In this paper we have only
discussed the incompleteness of the measured rates due to
the increasing fraction of SFD occurring in dusty environments. There are
other possible sources of incompleteness that could be present and could give
important contributions at high redshift. For example, a class of faint CC SNe
have been recently detected in the local universe (Pastorello et al. 2004)
that could be associated with 
explosion energy so small that most of the $^{56}$Ni falls back onto the
compact stellar remnant.
Such an effect was also invoked by Della Valle et al.  (2006) 
to explain the absence of any observable SN associated with the long 
GRB060614. These massive stars contribute to the far-IR luminosity of the
galaxies but not the observed CC SN rate.
The objects of this class seem to be quite rare in the local
universe, about 5\% of the total type II rate according 
to Pastorello et al. (2004), but 
the evolution of their rate with with redshift is not known. 
If they become more common at high redshifts, for
example due to metallicity effects, the expected rates will be lower than
presented here.

\bigskip

Summarizing, on the basis of the recent estimates of the density evolution of 
the star formation activity in clean and dusty environments, we have estimated
the fraction of SNe that are likely to be missed by the optical searches
up to z=2. We find a strong evolution of the missing fraction: 
for CC SNe it is as small
as 5-10\% in the local universe and rises to $\sim$30\% at z=1 and to
$\sim$60\% at z=2; 
for type Ia SN it is about 2\% at z=0, 15\% at z=1 and 35\% at z=2.
This hidden amount of SNe must be taken into account when
using the SN rate to put constraints on the cosmic SFD
and to design the future SN searches at high redshift.
\\

{\bf Acknowledgments}

We are grateful to A. Hopkins, R. Chary and P. P\'erez-Gonz\'alez
for useful discussions and for having provided useful data for Fig.~1.



\begin{thebibliography}{}
\bibitem[]{} 
  Babbedge, T. S., et al. 2006, MNRAS, 370, 1159
\bibitem[]{} 
  Baldry, I. K., Glazebrook, K. 2003, ApJ, 593, 258
\bibitem[]{} 
  Baldry, I. K., Glazebrook, K., Budav\'ari, T., et al., 2005, MNRAS, 358, 441
\bibitem[]{} 
  Barger, A. J., Cowie, L. L., Richards, E. A. 2000, AJ, 119, 2092
\bibitem[]{} 
  Bouwens, R. J. \& Illingworth, G. D., 2006, Nature, 443, 189
\bibitem[]{} 
  Burgarella, D., Perez-Gonzalez, P. G., Tyler, K. D. et al.
  2006, A\&A, 450, 69
\bibitem[]{} 
  Calura, F. \& Matteucci, F., 2006, ApJ, 652, 889
\bibitem[]{} 
  Cappellaro, E., Evans, R., \& Turatto, M., 1999, A\&A, 351, 459
\bibitem[]{} 
  Cappellaro, E., Riello, M., Altavilla, G., et al., 2005, A\&A, 430, 83
\bibitem[]{} 
  Chary, R., Dickinson, M. E., Tepliz, H. I., Pope, A., \& Ravindranath, S.,
  2005, ApJ, 635, 1022
\bibitem[]{} 
  Chary, R., et al., 2007, in ``At the Edge of the Universe'', 
  October 2006, Sintra, Portugal
\bibitem[]{} 
  Choi, P. I., Yan, L., Im, M., et al., 2006, ApJ, 637, 227
\bibitem[]{} 
  Condon, J. J. 1989, ApJ, 338, 13
\bibitem[]{} 
  Condon, J. J., Cotton, W. D., Broderick, J. J. 2002, AJ, 124, 675
\bibitem[]{} 
  Connolly, A. J., Szalay, A. S., Dickinson, M., SubbaRao, M. U.,
  Brunner, R. J. 1997, ApJL, 486, L11
\bibitem[]{} 
  Cowie, L. L., Songaila, A., Barger, A. J. 1999, AJ, 118, 603
\bibitem[]{} 
  Cram, L., Hopkins, A., Mobasher, B., \& Rowan-Robinson, M., 
  1998, ApJ, 507, 155
\bibitem[]{} 
  Cresci, G., Mannucci, F., Della Valle, M., \& Maiolino, R.
  2006, A\&A, 462, 927
\bibitem[]{} 
  Cresci, G., Mannucci, F., Maiolino, R., Della Valle, M., \& Ghinassi, F.	
  2002, IAUC, 7784, 1
\bibitem[]{} 
  Daddi, E., Dickinson, M., Chary, R., et al., 
  2005, ApJL, 631, L13
\bibitem[]{} 
  Dahlen, T., Strolger, L.-G., Riess, A., et al. 2004, ApJ, 613, 189
\bibitem[]{} 
  Dahlen, T., Mobasher, B., Dickinson, M., Ferguson, H. C.,
  Giavalisco, M., Kretchmer, C., \& Ravindranath, S., 
  2006, 654, 172
\bibitem[]{} 
  Della Valle, M., Panagia, N., Padovani, P., Cappellaro, E., 
  Mannucci, F., Turatto, M., 2005, ApJ, 629, 750
\bibitem[]{} 
  Della Valle, M., Chincarini, G., Panagia, N., et al., 
  2006, Nature, 444, 1050
\bibitem[]{} 
  Di Paola, A., Larionov, V., Arkharov, A., et al.,
  2002, A\&A, 393, L21
\bibitem[]{} 
  Farrara, A.., \& Ricotti, M., 2006, MNRAS, 373, 571
\bibitem[]{} 
  Filippenko, A. V., Li, W. D., Treffers, R. R., \& Modjaz, M. 
  2001, in ``Small Telescope Astronomy on Global Scales'', 
  ASP Conference Series Vol. 246, IAU Colloquium 183, B. Paczynski, 
  W.-P. Chen, and C. Lemme eds. 
  San Francisco: Astronomical Society of the Pacific, p. 121
\bibitem[]{} 
  Flores, H., et al. 1999, ApJ,  517, 148
\bibitem[]{} 
  Georgakakis, A., Hopkins, A. M., Sullivan, M., Afonso, J.,
   Georgantopoulos, I.,  Mobasher, B., Cram, L. 2003, MNARS, 345, 939
\bibitem[]{} 
  Greggio, L., 2005, A\&A, 441, 1055
\bibitem[]{} 
   Haarsma, D. B., Partridge, R. B., Windhorst, R. A., Richards, E. A.
   2000, ApJ, 544, 641
\bibitem[]{} 
   Kewlwy, L. J., Geller, M., J., Jansen, R., A., \& Dopita, M., A.,
   2002, AJ, 124, 3115
\bibitem[]{} 
  Hacking, P., Condon, J J., \& Houck, J. R., 1987, ApJL, 316, L15
\bibitem[]{} 
  Hopkins, A. M., \& Beacom, J., F., 2006, ApJ, 651, 142
\bibitem[]{} 
  Hopkins, A. M., Connolly, A. J., Haarsma, D. B., \& Cram, L. E.
  2001, ApJ, 122, 288
\bibitem[]{} 
  Jansen, R. A., Franx, M., \& Fabricant, D., 
  2001, ApJ, 551, 825
\bibitem[]{} 
  Le Floch, E.,  Papovich, C., Dole, H., et al.,
  2005, ApJ, 632, 169 
\bibitem[]{} 
  Li, W., 2005, 
  snap.lbl.gov/supernova\_workshop/snap2005.ppt
\bibitem[]{} 
  Lagache, G., Puget, J.-L., Dole, H., 
  2005, ARA\&A, 43, 727
\bibitem[]{} 
  Liang, Y. C., Hammer, F., Flores, H., Elbaz, D., Marcillac, D., and
  Cesarsky, C. J., 2004, A\&A, 423, 867
\bibitem[]{} 
  Lilly, S. J., Le F{\`e}vre, O., Hammer, F., Crampton, D.
  1996, ApJL, 460, L1
\bibitem[]{} 
  Lonsdale, C. J., Diamond, P. J., Thrall, H., Smith, H. E., Lonsdale, C. J.
  2006, ApJ, 647, 185
\bibitem[]{} 
  Machalski, J., Godlowski, W. 2000, A\&A, 360, 463
\bibitem[]{} 
  Madgwick, D., Hewett, P. C., Mortlock, D. J., \& Wang, L., 
  2003, ApJL, 599, L33
\bibitem[]{} 
  Maiolino, R., Della Valle, M., Vanzi, L., \& Mannucci, F.
  2001, IAUC, 7661, 2
\bibitem[]{} 
  Maiolino, R., Vanzi, L., Mannucci, F., Cresci, G., 
  Ghinassi, F., Della Valle, M.	
  2002, A\&A, 389, 84
\bibitem[]{} 
  Maiolino, R., Schneider, R., Oliva, E., Bianchi, S., Ferrara, A., 
  Mannucci, F., Pedani, M., \& Roca Sogorb, M.
  2004a, Nature, 431, 533
\bibitem[]{} 
  Maiolino, R., Oliva, E., Ghinassi, F., Pedani, M., Mannucci, F., 
  Mujica, R., \& Juarez, Y.
  2004b, A\&A 420, 889
\bibitem[]{} 
  Mannucci, f. \& Beckwith, S. V. W.,
  1995, ApJ, 442, 569
\bibitem[]{} 
  Mannucci, F., Maiolino, R., Cresci, G, et al.,
  2003, A\&A, 401, 519
\bibitem[]{} 
  Mannucci, F., Della Valle, M., Panagia, N., et al., 2005, A\&A, 433, 807 
\bibitem[]{} 
  Mannucci, F., Buttery, H., Maiolino, R., Marconi, A., \& Pozzetti, L.
  2007, A\&A, 461, 423
\bibitem[]{} 
  Mannucci, F., Della Valle, M., \& Panagia, N., 2006, MNRAS, 370, 773
\bibitem[]{} 
  Marcillac, D., Elbaz, D., Charlot, S., Liang, Y. C., Hammer, F., 
  Flores, H., Cesarsky, C., \& Pasquali, A.
  2006, A\&A, 458, 369
\bibitem[]{} 	
  Massarotti, M., Iovino, A., Buzzoni, A. 2001, ApJL, 559, L105
\bibitem[]{} 	
  Matteucci, F., \& Greggio, L., 1986, A\&A, 154, 279
\bibitem[]{} 	
  Matteucci, F., Panagia, N., Pipino, A., Mannucci, F., Recchi, S., 
  \& Della Valle, M.  2006, MNRAS, 372, 265
\bibitem[]{} 	
  Mattila, S., Meikle, W. P. S., \& Greimel, R.,
  2004, NewAR, 48, 595
\bibitem[]{} 
  Mauch, T. 2005, PhD Thesis, University of Sydney
\bibitem[]{} 
  Neill, J. D., Sullivan, M., Balam, D., et al., 2006, AJ, 132, 1126
\bibitem[]{} 
  Pain, R., Fabbro, S., Sullivan, M., et al., 2002, ApJ, 577, 120
\bibitem[]{} 
  P\'erez-Gonz\'alez, P. G., Rieke, G., Egami, E. et al., 2005, ApJ, 630, 82
\bibitem[]{} 
  P\'erez-Gonz\'alez, P. G., Zamorano, J., Gallego, J., Arag\'on-Salamanca, A.,
  \& Gil de Paz, A.
  2003, ApJ, 591, 827
\bibitem[]{} 
  Richmond, M. W., Filippenko, A. V., \& Galisky, J., 1998, PASP, 110, 553
\bibitem[]{} 
  Sadler, E.M., et al., 2002, MNRAS, 329, 227
\bibitem[]{} 
  Schiminovich, D., et al. 2005, APJL, 619, L47
\bibitem[]{} 
  Serjeant, S., Gruppioni, C., \& Oliver, S. 2002, MNRAS, 330, 621
\bibitem[]{} 
  Soifer, B. T., \& Neugebauer, G., 1991, AJ, 101, 354
\bibitem[]{} 
  Strolger, L.G., 2003, Phd thesis, University of Michigan
\bibitem[]{} 
  Sullivan, M., Treyer, M. A., Ellis, R. S., Bridges, T. J., Milliard, B.,
  Donas, J., 2000, MNRAS, 312, 442
\bibitem[]{} 
  Sullivan, M., Mobasher, B., Chan, B., Cram, L., Ellis, R., Treyer, M., 
  \& Hopkins, A.
  2001, ApJ, 558, 72
\bibitem[]{} 
  Thompson, R. I., Eisenstein, D., Fan, X., Dickinson, M., Illingworth, G.,
   Kennicutt, R. C. 2006, ApJ, 647, 787
\bibitem[]{} 
  Treyer, M. A., Ellis, R. S., Milliard, B., Donas, J., Bridges, T. J.
  1998, MNRAS, 300, 303
\bibitem[]{} 
  Valentini, G., Di Carlo, E., Massi, F.,  et al.
  2003, ApJ, 595, 779
\bibitem[]{} 
  van Buren, D., \& Norman, C. A., 1989, ApJ, 336, L67
\bibitem[]{} 
  Vijh, U. P., Witt, A. N., \& Gordon, K., D., 2003, ApJ, 587, 533
\bibitem[]{} 
  Wilson, G., Cowie, L. L., Barger, A., Burke, D. J. 2002, AJ, 124, 1258
\bibitem[]{} 
  Woosley, S. E. \& Weaver, T. A., 1986, ARA\&A, 24, 205
\bibitem[]{} 
  Wyder, T. K., et al., 2005, ApJL, 619, L15
\bibitem[]{} 
   Wolf, C., Meisenheimer, K., Rix, H.-W., Borch, A., Dye, S.,
   Kleinheinrich, M. 2003, A\&A, 2003, 401, 73
\end{thebibliography}
\end{document}